\documentclass[12pt]{article}

\usepackage{graphicx}
\usepackage{color}
\usepackage{graphicx}
\usepackage{dcolumn}
\usepackage{bm}
\usepackage{amsmath}
\usepackage{amsfonts}
\usepackage{setspace}
\usepackage{feynmp}

\def\simge{\mathrel{%
   \rlap{\raise 0.511ex \hbox{$>$}}{\lower 0.511ex \hbox{$\sim$}}}}

\def\simle{\mathrel{
   \rlap{\raise 0.511ex \hbox{$<$}}{\lower 0.511ex \hbox{$\sim$}}}}

\def\s#1{\setbox0=\hbox{$#1$}%
\rlap{\ifdim\wd0>.7em\kern.22\wd0\else\kern.1\wd0\fi /}#1}

\newcommand{\matel}[3]{\langle #1|#2|#3\rangle}
\newcommand{\vev}[1]{\langle #1 \rangle}

\newcommand{\ym}{y_m}

\begin{document}

\begin{titlepage}
\begin{flushright}
\begin{tabular}{l}
 SHEP
    28-10 \\ CP3-Origins-2010-38
\end{tabular}
\end{flushright}

\vskip1.5cm
\begin{center}
  {\Large \bf \boldmath  Scaling relations for the entire spectrum in
  mass-deformed conformal gauge theories} 
  \vskip1.3cm 
  {\sc Luigi Del Debbio$^{\,a}$\footnote{luigi.del.debbio@ed.ac.uk}  \&
    Roman Zwicky$^{\,b}$\footnote{Roman.Zwicky@soton.ac.uk}}
  \vskip0.5cm
  
  $^a$ {\sl School of Physics and Astronomy, University of Edinburgh, 
    Edinburgh EH9 3JZ, Scotland} \\
  $^b$ {\sl School of Physics \& Astronomy, University of Southampton, 
    Highfield, Southampton SO17 1BJ, UK} \, 
  \vspace*{1.5mm}
\end{center}

\begin{abstract}
We consider mass-deformed conformal gauge theories (mCGT) and
investigate the scaling behaviour of hadronic observables as a
function of the fermion mass.  Applying renormalization group
arguments directly to matrix elements, we find $m_H \sim
m^{1/(1+\gamma_*)}$ and $F \sim m^{\eta_F(\gamma_*)}$ for given $\eta_F(\gamma_*)$, for the 
hadronic masses and the decay
constants respectively, thereby generalizing our results from a previous paper to
the entire spectrum.  Applying the Hellmann-Feynman theorem to
the trace anomaly we obtain the hadron mass scaling independent of renormalization group
arguments. From the trace anomaly we obtain a relation reminiscent
of the Gell-Mann Oakes Renner relation in QCD.
Using the new results we  discuss the scaling of the
$S$-parameter inside the conformal window. 
Finally, we discuss how spectral representations can be
used to relate the mass and decay constant trajectories.
\end{abstract}

\end{titlepage}

\title{\Large }

\section{Introduction}

The physics of quantum field theories with an infrared fixed point
(IRFP) is characterized by scale invariance at large distances, which
implies a massless spectrum, and unbroken chiral symmetry. 
Gauge theories coupled to massless fermions that remain
asymptotically free in the ultraviolet and have an IRFP are said to
lie in the {\em conformal window}.  An example of fixed-point, in the
weak coupling regime, is due to Banks and Zaks~\cite{Banks:1981nn}.
For lower values of $N_f$ the critical coupling is expected to be
larger. In supersymmetry, where the nonperturbative dynamics can be
studied by analytical methods - see
e.g. Ref.~\cite{Intriligator:1995au} for a review - there are examples
of strong coupling fixed-points by virtue of electric-magnetic
duality.

Identifiying a conformal fixed point at strong coupling in a
nonsupersymmetric theory is an interesting theoretical challenge. 
Techniques have been developed in recent years to understand
the phase structure of gauge theories, see
e.g. Refs.~\cite{Sannino:2009za,Poppitz:2009uq} for recent results and
references.  In this work we will consider gauge theories minimally
coupled to $N_f$ Dirac fermions in arbitrary representations of the
gauge group. All fermions are degenerate, even though different
fermion masses can be easily accomodated. The fermionic mass term is a
relevant deformation that drives the theory away from conformality. We
refer to these theories as mass-deformed conformal gauge theories
(mCGT). 
 Mass deformed theories develop a string tension and thus
lead to the formation of massive bound states which shall call hadrons.~\cite{Miransky:1998dh,DelDebbio:2009fd}
It is the scaling of the hadronic
spectrum and the decay constants as a function of the fermion mass
and its applications that we want to consider in this study.

Besides their theoretical interest, conformal theories deformed away
from the fixed point are interesting candidates technicolor~\cite{Sannino:2004qp,Luty:2004ye,Sannino:2008nv} model building. The important issue for
phenomenology is the characterization of the fixed point, through the
computation of the relevant critical exponents. 
The latter are relevant for technicolor searches at
colliders~\cite{Galloway:2010bp,Chivukula:2010tn}.

The lattice formulation of gauge theories provides a  powerful tool to
investigate the nonperturbative dynamics from first principles. Monte
Carlo simulations are necessarily performed at finite fermion mass,
the signature of an IRFP can be found by studying the scaling of
physical quantities with the fermion mass as proposed in
Refs.~\cite{Luty:2008vs}. The scaling laws are dictated
by the critical exponents of the fixed point, and have been studied
intensively in recent numerical
investigations~\cite{DelDebbio:2009fd,DeGrand:2009hu,DelDebbio:2010hu}.

In this work we:
\begin{itemize}
\item provide scaling laws for condensates, hadron masses, decay constants for the \emph{entire} spectrum.
\item derive the scaling law for the hadron masses from the trace anomaly and the Hellmann-Feynman theorem 
without using RG arguments. The relation obtained from 
the trace anomaly  is reminiscent of the GMOR relation in QCD.
\item reiterate on scaling laws for the S-parameter using our results.
\item give the relation between  the mass and decay constant trajectories
\item suggest to probe the beta function via the trace anomaly 
and provide a presentation of the beta function which resembles the NSVZ beta-function \cite{NSVZ}.
\end{itemize}

In connection with the first item it is worthwhile to note out that 
in a previous work~\cite{DelDebbio:2010ze}, we have derived the
scaling of the hadron masses and decay constants, for the lowest-lying state in any channel with given quantum
numbers, using renormalization
group (RG) arguments on correlators for large Euclidian times.

\section{Mass scaling from anomalous dimensions} 

Consider an operator ${\cal O}(x;\mu)$ renormalized at the scale
$\mu$, with scaling dimension $\Delta_{\cal O}$ and quantum numbers
such that the matrix element 
\begin{equation}
  \label{eq:matel}
  T_{\varphi_1 {\cal O} \varphi_2}(g, \hat m, \mu) \equiv  
  \matel{\varphi_2}{{\cal O}(0)}{\varphi_1}   \;, \quad \Delta_{\cal O} =
  d_{\cal O} + \gamma_{\cal O} \;
\end{equation}
is non-vanishing.  The symbols $d_{\cal O}$ and $\gamma_{\cal O}$ denote the engineering and anomalous dimension 
respectively. We have indicated the dependence on the two
renormalized parameters of the mCGT -- the rescaled mass $\hat m \mu =
m$, and the gauge coupling $g$ -- as well as on the scale $\mu$.  
The
kets $|\varphi_{i} \rangle$, for $i = 1,2$, denote physical states
with scaling dimensions $\Delta_{\varphi_i} = d_{\varphi_i}$.

We closely follow part of the argumentation in our previous
paper~\cite{DelDebbio:2010ze}, where we applied renormalization group
(RG) arguments to study the scaling of field correlators; here we
focus on matrix elements of the type introduced in
Eq.~(\ref{eq:matel}), adding to the discussion the observation that
the physical states are free of anomalous dimensions  (i.e. 
$\Delta_{\varphi_i} = d_{\varphi_i}$). Close to the
fixed point the RG transformation $\mu = b \mu'$ on the matrix element
results in:
\begin{equation}
  \label{eq:step1}
  T_{\varphi_1 {\cal O} \varphi_2}(g,\hat{m},\mu) = 
  b^{-\gamma_{\cal O}} 
  T_{\varphi_1 {\cal O} \varphi_2}(g^\prime,\hat{m}^\prime,\mu^\prime) \, .
\end{equation} 
Near a non-trivial fixed point the couplings display powerlike
behaviour:
\begin{equation}
  \label{eq:scal1}
  g^\prime = b^{y_g} g \, ,~~~~ \hat m^\prime = b^{y_m} \hat m\, ,
\end{equation}
with 
\begin{equation}
\label{eq:ym}
y_m = 1 + \gamma_* \;,
\end{equation}
where $\gamma_*$ denotes the anomalous dimension of the mass at the fixed point.
Neglecting  the irrelevant coupling $g$ ($y_g < 0$)\footnote{ We remind the reader that the assumption is, to be verified by lattice simulations, that we are studying the
theory in a neighbourhood of a fixed point where the mass is the only
relevant coupling.} and multiplying
all mass units by the factor $b$ we obtain:
\begin{equation}
  \label{eq:step2}
  T_{\varphi_1 {\cal O} \varphi_2}(\hat{m}^\prime,\mu^\prime) = 
  b^{- (d_{\cal O}  + d_{\varphi_1} + d_{\varphi_2}     )   }
  T_{\varphi_1 {\cal O} \varphi_2}(\hat{m}^\prime,\mu) \;,
\end{equation} 
Choosing $b$ such that $\hat m^\prime=1$, combining equations \eqref{eq:step1} and 
\eqref{eq:step2} yields:
\begin{equation}
  \label{eq:final}
  T_{\varphi_1 {\cal O} \varphi_2}(\hat{m},\mu) \sim  
  \left(  \hat m \right) ^{(\Delta_{\cal O} + d_{\varphi_1} + d_{\varphi_2})/y_m }   \, .
\end{equation}
Note that in the approximation made there are no higher order
corrections.  Such correction originate from taking into account
corrections to the beta function of the form $\beta = A
m^{\eta_\beta}$ with $\eta_\beta > 0$.

The result in Eq.~(\ref{eq:final}) can also be derived starting from
correlation functions. For example for the matrix element the operator
$\mathcal O$ between one physical state and the vacuum we have:
\begin{align}
  \label{eq:LSZ}
  (p_1^2 &- m_{\varphi_1}^2)  i  \int d^4\! x e^{i  p_1 \cdot  x}  
  \matel{0}{T {\cal O}(0) \Phi_1(x) } {0} = \nonumber \\    
  &= (p_1^2 - m_{\varphi_1  }^2)  \left(\frac{\matel{0}{{\cal
          O}}{\varphi_1} 
      \matel{\varphi_1}{\Phi_1}{0}}{p_1^2 -
      m_{\varphi_1  }^2}    
    + \dots  \right) \, ,
\end{align}
where $\Phi_1$ denotes an interpolating operator for the state
$\varphi_1$. All other contributions but the $\varphi_1$-pole
vanish for on-shell momenta.  Generalization to the case of two
physical states is straightforward.

The scaling relations for the decay constants given in Table 1 of
Ref.~\cite{DelDebbio:2010ze} can be readily rederived from the
considerations presented above, borrowing the relation 
$m_H \sim m^{1/(1+\gamma_*)}$ from the next section. 
They apply trivially to {\em all}
one-particle stable states in the spectrum and not only to the lowest lying state as in Ref.~\cite{DelDebbio:2010ze}.

\begin{table}[ht]
  \label{tab:tab_decay}
  \centering
  \begin{tabular}{l|l|l|l|l|r}
    $\cal O$ & {\rm def} & $\matel{0}{ {\cal O} }{ J^\mathrm{P(C)}(p)
    }$ & $J^\mathrm{P(C)}$ & $\Delta_{\cal O}$  & $\eta_{G[F]}$
    \\[0.1cm] 
    \hline 
    $S$ & $\bar qq$ & $G_{S}$ & $0^{++}$ & $3 - \gamma_*$ & 
    $(2-\gamma_*)/\ym$ \\[0.1cm] 
    $S^a$ & $\bar q \lambda^a q$ & $G_{S^a}$ & $0^{+}$ & $3 - \gamma_*$ &
    $(2-\gamma_*)/\ym$ \\[0.1cm] 
    $P^a$ & $\bar q i\gamma_5 q$ & $G_{P^a}$ & $0^{-}$ & $3 - \gamma_*$ &
    $(2-\gamma_*)/\ym$ \\[0.1cm] 
    $V$ & $\bar q \gamma_\mu q$ & $\epsilon_\mu(p) M_V F_{V} $ &
    $1^{--}$ & $3$ & $1/\ym$ \\[0.1cm]
    $V^a$ & $\bar q \gamma_\mu \lambda^a q$ & $\epsilon_\mu(p) M_V
    F_{V^a} $ & $1^{-}$ & $3 $ & $1/\ym$ \\[0.1cm] 
    $A^a$ & $\bar q \gamma_\mu \gamma_5 \lambda^a q$ & $\epsilon_\mu(p) M_A
    F_{A^a}  $ & $1^{+}$ & $3$ & $1/\ym $   \\[0.1cm] 
    &  & $i p_\mu F_{P^a}$ & $0^{-}$ & $3$ & $1/\ym$  
  \end{tabular} 
  \caption{\small Scaling laws, $G[F]\sim m^{\eta_{G[F]}}$ for decay constants.
    The symbol $\ym \equiv 1 + \gamma_*$ denotes the
    scaling dimension of the mass 
    and $\Delta_{\cal O} = d_{\cal O} + \gamma_{\cal O}$. 
    The symbol $a$ denotes the adjoint flavour index, and $\lambda^a$
    are the generators normalized as $\mathrm{tr}[\lambda^a \lambda^b]
    =2 \delta^{ab}$. No such simple expression exists for the axial singlet current because
    of the chiral anomaly \cite{DelDebbio:2010ze}.}  
 \end{table}

\subsection{Scaling of hadron masses}
The approach that we have introduced in the previous section, namely
the study of matrix elements of given operators between physical
states, can also be used to investigate the scaling of the hadron
masses with the fermion mass.

RG arguments applied to two-point functions in
Refs.~\cite{DelDebbio:2010ze}\cite{DelDebbio:2009fd}\footnote{ This relation was first proposed in 
Ref.~\cite{Miransky:1998dh} by looking at the pole mass around the Banks-Zaks type fixed-point.}  led to the scaling relation:
\begin{equation}
  \label{eq:MH}
  M_H \sim m^{1/y_m} \, ,
\end{equation}
for the lowest state in any given channel. Here we shall generalize it
to the entire spectrum using the trace or scale anomaly of the energy
momentum tensor.

Let us first state two general facts.  When considering its matrix
elements between on-shell (i.e. physical) states 
, the trace of
the energy momentum tensor assumes the following form~\cite{EMTtrace}\footnote{When evaluated on states, in this notation, it is understood that only the connected part is evaluated.}:
\begin{equation}
  \label{eq:trace}
  \theta_\alpha^{\phantom{x}\alpha}|_{\rm on-shell} = \frac{1}{2 g}{\beta} G^2 + N_f m (1
  + \gamma_m)  \bar q q  \, ,
\end{equation}
where $\beta = \frac{\partial{ g}}{d \ln \mu}$. 
On the other
hand the matrix element of the energy momentum tensor between two
physical states can be written as, e.g. \cite{DGH},:
\begin{equation}
  \label{eq:decomp}
  \matel{H(p)}{\theta_{\alpha \delta}}{H(p)} = 2 p_\alpha p_\delta  \;,
\end{equation}
which is consistent with the relativistic normalization $\vev{H(\vec{p})|H(\vec{k})
} = 2 E_{p} \delta^{(3)}(\vec{p}- \vec{k})$. Note that all
operators appearing in eq.~(\ref{eq:decomp}) are intended to be
renormalized at some scale $\mu$. 

Taking the trace of Eq.~(\ref{eq:decomp}), equating with
(\ref{eq:trace}), neglecting the $\beta$ function, and adopting our
notation $\gamma_m = \gamma_*$ yields
\begin{equation}
  \label{eq:GMOR-like}
  2 M_H^2 = N_f (1+ \gamma_*) m \matel{H}{\bar q q}{H}\, ,
\end{equation}
a relation reminiscent of the Gell-Mann Oakes Renner relation in QCD.
The scaling of the hadron masses, for the entire spectrum, can be obtained in two
alternative ways.
\begin{enumerate}
\item From \eqref{eq:final}, with $d_{H(p)} = -1$ and $\Delta_{m\bar
    qq} = 4$, it follows that $m \matel{H}{\bar q q}{H} \sim
  m^{2/y_m}$ and thus Eq.~(\ref{eq:GMOR-like}) implies (\ref{eq:MH}).
  For the scaling corrections, the same remarks apply as for the
  quantity in Eq.~(\ref{eq:final}).
\item The Feynman-Hellmann  theorem states that,
  \begin{equation}
    \label{eq:FH}
    \frac{\partial E_\lambda}{\partial \lambda}  = 
    \matel{\psi(\lambda)} {\frac{\partial \hat{H}(\lambda)}
      {\partial \lambda}}{\psi(\lambda)} \, ;
  \end{equation}
  the variation of the energy with respect to a parameter equals the
  expectation value of the variation of the Hamiltonian. In our case, we
  consider the derivative with respect to the mass $m$. Taking into
  account our chosen normalization of states, Eq.~(\ref{eq:FH}) yields
  \begin{equation}
    m \frac{\partial M_H^2}{\partial m} = N_f m \matel{H}{\bar q q}{H}\, .
  \end{equation}
  Equating with Eq.~(\ref{eq:GMOR-like}) we get 
  \begin{equation}
    m \frac{\partial M_H}{\partial m} = \frac{1}{1+\gamma_*} M_H \, ,
  \end{equation}
  which implies Eq.~(\ref{eq:MH}) with \eqref{eq:ym}. It is worth emphasizing that this derivation does
  \emph{not} depend on RG arguments.
\end{enumerate}

Note that the same statement applies to the decay rates $\Gamma_{A \to
  BC..}  \sim m^{1/y_m}$ since the latter transition Hamiltonian is
free from anomalous scaling. Similar arguments for the scaling of
decay rates were already discussed in our previous
paper~\cite{DelDebbio:2010ze}.

\subsection{A remark on the $\beta$-function}

Finally, Eqs.~(\ref{eq:trace}) and~(\ref{eq:decomp}) imply a
relation between the beta function and the mass anomalous
dimension:
\begin{equation}
  \label{eq:exbeta}
  \beta = \frac{A_H + \gamma_m B_H}{G_H}\, ,
\end{equation}
where
\begin{align}
  A_H &= 2 M_H^2 - m N_f \langle H | \bar q q | H \rangle \, , \nonumber \\
  B_H &= m N_f \langle H | \bar q q | H \rangle \, , \nonumber \\
  G_H &= \langle H | G^2 | H \rangle \, . \nonumber
\end{align}
The detailed investigation of this relation is postponed to further
studies \cite{prep}. The connection with the so-called NSVZ beta function \cite{NSVZ} is obvious.  
This relation could, for instance, be used to probe the $\beta$-function 
in mCGT. Note that the combinations $A_H/G_H$ and $B_H/G_H$ are independent of the state $H$. 

\subsection{Remarks on S-parameter}
 The determination of the S-parameter for gauge theories is
  important for technicolor model building, be it for walking-like or
  conformal behaviour.  In this section, we discuss the implications
of our scaling relations for the S-parameter. These analytical results
for the scaling of the S-parameter lead to new criteria that can help
to distinguish a conformal phase from a walking phase.

Using the results valid for the entire spectrum rather than the lowest
lying state, we can put the remarks on the $S$ parameter of mCGT that
were presented in the conclusions of Ref.~\cite{DelDebbio:2010ze} on
more solid grounds.

Defining the dimensionless correlation function $\Pi_{V\!-\!A}(q^2)$
from the correlation function:
\begin{align}
  \Pi(q)^{\mu \nu}_{ab}  &\equiv  i \int \! d^4 \!x e^{i q \cdot  x} 
  \matel{0}{T\left( V^\mu_a(x)V^\nu_b(0) - (V \leftrightarrow A)
    \right) }{0}  \\
  &=
  (- q^2 g^{\mu\nu}  + q^{\mu}q^{\nu}) \delta_{ab} \Pi_{V\!-\!A}(q^2) +  g_{\mu\nu} m^2 \Pi_{A}(q^2)  \, . 
  \nonumber
\end{align}
Saturating the correlation function with hadronic states, one
would expect
\begin{equation}
  \label{eq:0}
  \Pi_{V\!-\!A}(q^2) = \frac{f_V^2}{m_V^2-q^2} - \frac{f_A^2}{m_A^2-q^2} -
  \frac{f_P^2}{ m_P^2-q^2} + ...    \, .
\end{equation}
As discussed in Ref.~\cite{Sannino:2010ca}, the scaling of the
correlator with the fermion mass depends on the order in which the
limits $q^2\to 0$ and $m\to 0$ are taken. 
When $q^2\to 0$ at fixed $m$, the scaling laws for the hadronic masses
and decay constants imply that 
\begin{equation}
  \label{eq:q2=0}
  \Pi_{V\!-\!A}(0)  \sim O(m^0)\, ;
\end{equation}
our analysis only yields the scaling law, but no information on the
prefactor. The analysis presented
in Refs.~\cite{Sannino:2010ca,Sannino:2010fh,DiChiara:2010xb} actually
aims at a lower bound for the prefactor. 
We would like to add that we cannot exclude that the vector and axial contributions conspire to cancel up some order in $m$.
For the case where we take the limit $m\to 0$ at fixed $q^2$, we obtain:
\begin{equation}
  \label{eq:01}
  \Pi_{V\!-\!A}(q^2) \stackrel{-q^2 \gg (\Lambda_{U})^2}{\sim}
  \frac{m^{2/y_m}}{q^2} \, 
  + O\left( \frac{m^2}{q^2} \right) \;,
\end{equation}
with the same caveat we highlighted for Eq.~(\ref{eq:q2=0}).  The
condition $-q^2 \gg (\Lambda_{\rm U})^2$, where $\Lambda_{\rm U}$ is
the scale where asymptotic freedom sets in, as otherwise the highest resonances
$M_{\rm high}^2 > q^2$ and this would imply a  behaviour as in Eq.~\eqref{eq:q2=0}.  
Note that for $-q^2 \gg (\Lambda_{\rm U})^2$ the correlator
$\Pi_{V\!-\!A}(q^2)$ does still receive perturbative correction of the
order $m^2/q^2$ due to the explicit breaking of chiral symmetry. These
corrections can already be seen in a perturbative computation near the
Banks-Zaks fixed point, as was pointed out in
Ref.~\cite{Sannino:2010ca}. As emphasized in
Ref.~\cite{DelDebbio:2010ze} the behaviour \eqref{eq:q2=0} is distinct
from the correlator of a gauge theory with broken chiral symmetry has
a pion pole and thus behaves like $\Pi(0) \sim O(m^{-1})$. This limit
is useful to identify a conformal behaviour in a lattice simulation,
where the limit $m\to 0$ at fixed $q^2$ can be
investigated~\cite{Shintani:2008qe,Boyle:2009xi,DeGrand:2010tm}.
 
\section{Relating the mass and decay constant trajectory}
We have established the scaling of hadronic masses and decay constants
in terms of the mass.  The nature of the spectrum remains an open
question, but we we can say something about the relation between the
masses and the decay constants in the large-$N_c$ limit.  In the latter
limit the width it is believed to be $N_c$ using the same 
$N_c$ scaling arguments as in QCD.
A two-point function, of an
operator ${\cal O}$ coupling to states $|H_n \rangle$, assumes the
following form:
\begin{equation}
  \Delta(q^2) \sim \int_x e^{ix\! \cdot \! q}\matel{0}{{\cal O}(x) {\cal O}(0)}{0} =
  \sum_n \frac{|g_{H_n}|^2} {q^2 + M_{H_n}^2} \;,
\end{equation}
in Euclidean space.  The symbol $g_{H_n} \equiv \matel{0}{{\cal
  O}}{H_n}$ defines what we call decay constant in this context. 

On the other hand when the mass is sent to zero, the spectral
representation for the two point function combined with scale
invariance implies~\cite{Stephanov:2007ry}:
\begin{equation}
  \label{eq:propi}
  \Delta(q^2) = \int_0^{\infty} \frac{ ds \, s^{1-\gamma_*}}{q^2+s}  
  + {\rm s.t}  \;\;    \sim  \;\; (q^2)^{1-\gamma_*} \;,
\end{equation}
where we have implicitly parametrized the scaling dimension of 
${\cal O}$ as $\Delta_{{\cal O}} = 3 - \gamma_*$.  The
symbol ${\rm s.t.}$ stands for ultraviolet subtraction terms which
are not relevant for our argument.  Thus the masses and decay
constants have to behave in such a way as to reproduce
\eqref{eq:propi}. It is readily seen that this is achieved as follows:
\begin{equation}
  M_{H_n}^2 \sim \alpha_n   m^{\frac{2}{1+\gamma_*}} \;, \quad 
  g_{H_n}^2 \sim  \alpha'_n (\alpha_n)^{1-\gamma_*}  m^{\frac{2(2-\gamma_*)}{1+\gamma_*}}
\end{equation}
The symbol $\alpha_n$ denotes any monotonic increasing function and
$\alpha'_n$ is the derivative w.r.t. to $n$.  The freedom of choosing
$\alpha_n$ corresponds to the freedom of changing variables in the
integral representation \eqref{eq:propi}\footnote{Similar considerations can 
also be applied to  large-$N_c$ QCD.
In the latter case a linear $\alpha_n \sim n$ (Regge trajectory)
spectrum is expected.
In the case where  
${\cal O} \to \bar q \gamma_\mu q$ and at large momentum transfer
$\Delta_{\mu\nu} \sim (q^2 g_{\mu\nu} - q_\mu q_\nu) \ln ( q^2)$ which is easily verified 
by perturbation theory justified by asymptotic freedom. 
This  suggests $\gamma^*|_{\rm eff} = 1$
and $g_n \sim O(n^0)$.
See e.g. Ref.~\cite{duality} where this type of model has been suggested. N.B. the role of 
the parameter $m$ is played by $\Lambda_{\rm QCD}$ and the powers are simply the engineering
dimensions of the mass and decay constants.}\footnote{Our discussion formally resembles
the deconstruction of a scale invariant spectrum in \cite{Stephanov:2007ry}. The difference
is that we give a specific interpretation of the spacing of the resonances whereas 
in \cite{Stephanov:2007ry} the spacing is introduced as a mathematical tool.}.

{\bf Acknowledgements:} LDD and RZ  acknowledge the support of advanced 
STFC fellowships. We are grateful to John Donoghue for correspondance and Nick Evans \& Tim Morris 
for discussions.

\end{document}